\documentclass[twocolumn]{aastex631}
\newcommand{\roma}[1]{\uppercase\expandafter{\romannumeral#1}}
\newcommand{\speed}[1]{#1 km~s${}^{-1}$}

\newcommand{\nfig}[1]{Figure~\ref{#1}}

\usepackage{amsmath}
\usepackage{subfigure}
\usepackage{hyperref}%

 \usepackage{url}
\usepackage{graphicx}
\usepackage{natbib}
\usepackage{amssymb,txfonts}
\usepackage{multirow}
\usepackage{array}
\usepackage{sidecap}
\usepackage{hyperref}
\usepackage{epstopdf}
\usepackage{footnote}
\usepackage{tabularx}
\usepackage{booktabs}

\hypersetup{
    bookmarks=true,         
    unicode=false,          
    pdftoolbar=true,        
    pdfmenubar=true,        
    pdffitwindow=false,     
    pdfstartview={FitH},    
    pdftitle={My title},    
    pdfauthor={Author},     
    pdfsubject={Subject},   
    pdfcreator={Creator},   
    pdfproducer={Producer}, 
    pdfkeywords={keyword1} {key2} {key3}, 
    pdfnewwindow=true,      
    colorlinks=true,        
    linkcolor=blue,         
    citecolor=blue,         
    filecolor=blue
    urlcolor=blue
    }

\accepted{April 7, 2022}
\shorttitle{Observations of a Flare-ignited broad Quasi-periodic Fast-propagating wave train}
\shortauthors{Zhou et al.}

\newcommand{\p}[1]{{\color{red}{#1}}}

\begin{document}
\title{Observations of a Flare-ignited broad Quasi-periodic Fast-propagating wave train}
\correspondingauthor{Yuandeng Shen}
\email{ydshen@ynao.ac.cn}
\author[0000-0001-9374-4380]{Xinping Zhou}
\affiliation{Yunnan Observatories, Chinese Academy of Sciences,  Kunming, 650216, China}
\affiliation{State Key Laboratory of Space Weather, Chinese Academy of Sciences, Beijing 100190, China}
\affiliation{University of Chinese Academy of Sciences, Beijing, China}
\author[0000-0001-9493-4418]{Yuandeng Shen}
\affiliation{Yunnan Observatories, Chinese Academy of Sciences,  Kunming, 650216, China}
\affiliation{State Key Laboratory of Space Weather, Chinese Academy of Sciences, Beijing 100190, China}
\affiliation{University of Chinese Academy of Sciences, Beijing, China}
\author{Ying D. Liu}
\affiliation{State Key Laboratory of Space Weather, Chinese Academy of Sciences, Beijing 100190, China}
\affiliation{University of Chinese Academy of Sciences, Beijing, China}
\author{Huidong Hu}
\affiliation{State Key Laboratory of Space Weather, Chinese Academy of Sciences, Beijing 100190, China}
\author{Jiangtao Su}
\affiliation{CAS Key Laboratory of Solar Activity, National Astronomical Observatories, Beijing 100012, China}
\author[0000-0003-0880-9616]{Zehao Tang}
\affiliation{Yunnan Observatories, Chinese Academy of Sciences,  Kunming, 650216, China}
\affiliation{University of Chinese Academy of Sciences, Beijing, China}
\author{Chengrui Zhou}
\affiliation{Yunnan Observatories, Chinese Academy of Sciences,  Kunming, 650216, China}
\affiliation{University of Chinese Academy of Sciences, Beijing, China}
\author[0000-0001-9491-699X]{Yadan Duan}
\affiliation{Yunnan Observatories, Chinese Academy of Sciences,  Kunming, 650216, China}
\affiliation{University of Chinese Academy of Sciences, Beijing, China}
\author{Song Tan}
\affiliation{Yunnan Observatories, Chinese Academy of Sciences,  Kunming, 650216, China}
\affiliation{University of Chinese Academy of Sciences, Beijing, China}

\begin{abstract}
Large-scale Extreme-ultraviolet (EUV) waves are frequently observed as an accompanying phenomenon of flares and coronal mass ejections (CMEs). Previous studies mainly focus on EUV waves with single wavefronts that are generally thought to be driven by the lateral expansion of CMEs. Using high spatio-temporal resolution multi-angle imaging observations taken by the Solar Dynamic Observatory and the Solar Terrestrial Relations Observatory, we present the observation of a broad quasi-periodic fast propagating (QFP) wave train composed of multiple wavefronts along the solar surface during the rising phase of a GOES M3.5 flare on 2011 February 24. The wave train transmitted through a lunate coronal hole (CH) with a speed of $\sim$840$\pm$\speed{67}, and the wavefronts showed an intriguing refraction effect when they passed through the boundaries of the CH. Due to the lunate shape of the CH, the transmitted wavefronts from the north and south arms of the CH started to approach each other and finally collided, leading to the significant intensity enhancement at the collision site. This enhancement might hint the occurrence of interference between the two transmitted wave trains. The estimated magnetosonic Mach number of the wave train is about 1.13, which indicates that the observed wave train was a weak shock. Period analysis reveals that the period of wave train was $\sim$90 seconds, in good agreement with that of the accompanying flare. Based on our analysis results, we conclude that the broad QFP wave train was a large-amplitude fast-mode magnetosonic wave or a weak shock driven by some non-linear energy release processes in the accompanying flare.
\end{abstract}

\section{Introduction}
Large-scale propagating wave-like disturbances at fast speeds of \speed{200-1500} in the solar corona \citep{2013ApJ...776...58N} were firstly observed by the Extreme Ultraviolet (EUV) Imaging Telescope \citep[EIT:][]{1995SoPh..162..291D} onboard the Solar and Heliospheric Observatory \citep[SOHO:][]{1995SoPh..162....1D}, and they were dubbed as EIT or EUV waves in history. During the past two decades, a mass of observational and theoretical studies have been performed to study the origin and the physical nature of the EUV waves, and these results indicate that EUV waves could either be explained as fast-mode shock/magnetohydrodynamic (MHD) waves or non-waves {caused} by the reconfiguration of coronal magnetic fields \citep[e.g.,][]{2008SoPh..253..215V,2014SoPh..289.3233L}. However, no single interpretation can satisfy all constraints imposed by the observations \citep{ 2017SoPh..292....7L}. In order to reconcile the observations, \cite{2002ApJ...572L..99C} predicted that there are two types of EUV waves in a solar eruption: a preceding fast-mode shock and a slower wave-like density perturbation caused by the stretching of magnetic field lines. So far, this scenario has been confirmed by many observations \citep[e.g.,][]{2011ApJ...732L..20C,2012ApJ...752L..23S,2014ApJ...795..130S}

Generally, flares and coronal mass ejections (CMEs) are spectacular phenomena that can potentially launch large-scale EUV waves. Thus, there are two main views on the generation of EUV waves. Some researchers favor that EUV waves are generated by the lateral expansion of the associated CMEs \citep[e.g.,][]{2002ApJ...572L..99C,2010ApJ...724L.188P,2012ApJ...754....7S,2019ApJ...871....8L,2021ApJ...911..118D,2022arXiv220213051H}. In this view, an EUV wave is generated by the combination of a piston-shock and a bow-shock owing to the expansion of a CME. Others prefer that the excitation of EUV waves is due to the pressure pulses produced by the accompanying flares \citep[e.g.,][]{2002A&A...383.1018K, 2004A&A...418.1117W, 2008SoPh..253..305M, 2016ApJ...832..128C, 2016ApJ...828...28K}. In this view, an EUV wave is driven by the flare-volume expansion caused by the impulsive energy release in a flare. Despite a mass of observational and numerical studies that have been performed to support the CME-driven scenario, believable evidence for supporting the flare-driven scenario is still scarce \citep{2008SoPh..253..215V}.
It should be pointed out here that some non-CME-association EUV waves are also not driven by flare pulses. For example, they can be driven by the fast expansion of lower coronal loops associated with failed solar eruptions \citep[e.g.,][]{2017ApJ...851..101S,2020ApJ...894..139Z}, or sudden loop expansion caused by remote eruptions \citep[e.g.,][]{2018ApJ...860L...8S}, or coronal jets \citep[e.g.,][]{2018ApJ...861..105S}.
Therefore, \textcolor{black}{ in the case when an EUV wave is not associated with a CME, one cannot conclude that this EUV wave must be driven by a flare pulse, while other physical mechanisms could still be possible.} Whereas one can check the eruption details with high-resolution imaging observations to clarify the truly driven mechanism of the EUV wave.

Believable evidence for EUV waves driven by flare pulses have been observed by Atmospheric Imaging Assembly \citep[AIA;][]{2012SoPh..275...17L} onboard Solar Dynamics Observatory \citep[SDO;][]{2012SoPh..275....3P}, which are seen as 
relatively small-scale wave trains along coronal loops and have similar periods with
their accompanying flares \citep[e.g.,][]{2011ApJ...736L..13L,2012ApJ...753...53S,2013SoPh..288..585S,2018ApJ...853....1S,2021arXiv211215098Z}. Large-scale quasi-periodic EUV wave trains, similar to typical single pulsed EUV waves, were also observed ahead of the CME bubble \citep[e.g.,][]{2012ApJ...753...52L}. However, the excitation mechanism of these EUV wave trains is still unclear. For example, in \cite{2012ApJ...753...52L} the wave trains has a common 2 minutes period with the
accompanying flare, while in \cite{2019ApJ...873...22S} the period of the wave trains 
 showed a large difference \textcolor{black}{from} that of the accompanying flare. In terms of intuition, such large-scale EUV wave trains are composed of multiple concentric wavefronts, and they are unlike to be driven by the expansion of CME bubbles. Therefore, \cite{2012ApJ...753...52L} proposed that the EUV wave train was possibly driven by the flare pulse since its period
was similar to the flare. In \cite{2019ApJ...873...22S}, since the period of the wave train was similar to the unwinding filament threads in the eruption source region, the authors alternatively proposed that the wave train was excited by the sequentially outward expansion of the unwinding filament threads. As for QFP wave trains, \cite{2021arXiv211214959S} divided them into narrow and broad QFP types based on their different physical properties. The former is characterized as propagating coherent wavefronts along the coronal loops with a relatively narrow angular width and a small intensity amplitude \citep{2011ApJ...736L..13L,2012ApJ...753...53S,zhou2022soph,2022arXiv220108982D}, while the latter propagates along the solar surface with a broad angular width and a relatively large intensity amplitude \citep{2019ApJ...873...22S}. In comparison, the physical parameters of broad QFP wave trains are more similar to the typical single pulsed EUV waves. The generation mechanism of broad QFP wave is still an open question \citep{2021arXiv211214959S}, although several numerical works have been performed \citep{2015ApJ...800..111Y,2016ApJ...823..150T,2017ApJ...847L..21P,2021ApJ...911L...8W}.

In this letter, we present observations of a broad QFP wave train propagating along the solar surface, whose period was similar to the accompanying flare's quasi-periodic pulsations (QPPs). The current event might provide a reliable case for supporting the flare-driven mechanism of the EUV waves. In addition, this study also provides the first evidence of the interference effect of EUV waves, suggesting the true wave nature of the observed disturbance. 

\section{RESULTS}
The broad QFP wave train was intimately associated with a partial halo CME and a GOES M3.5 flare. The CME had an average speed of $\sim$\speed{1186} \footnote[5]{\url{https://cdaw.gsfc.nasa.gov/CME\_list/} }, and the flare's start and peak times were at 07:23 UT and 07:35 UT \footnote[6]{\url{https://hesperia.gsfc.nasa.gov/goes/goes\_event\_listings/}}, respectively. Although the wave train can be identified in all EUV channels of AIA, we concentrate principally on the AIA 171 \AA, 193 \AA, and 211 \AA\ channels to obtain essential details of the eruption in this study. In addition, the soft and hard X-ray fluxes recorded respectively by the GOES and the {\em Reuven Ramaty High Energy Solar Spectroscopic Imager} \citep[RHESSI;][]{2002SoPh..210....3L} are also used to analyze the periodicity of the flare QPP. 

\begin{figure*}
\epsscale{1.}
\begin{center}
\plotone{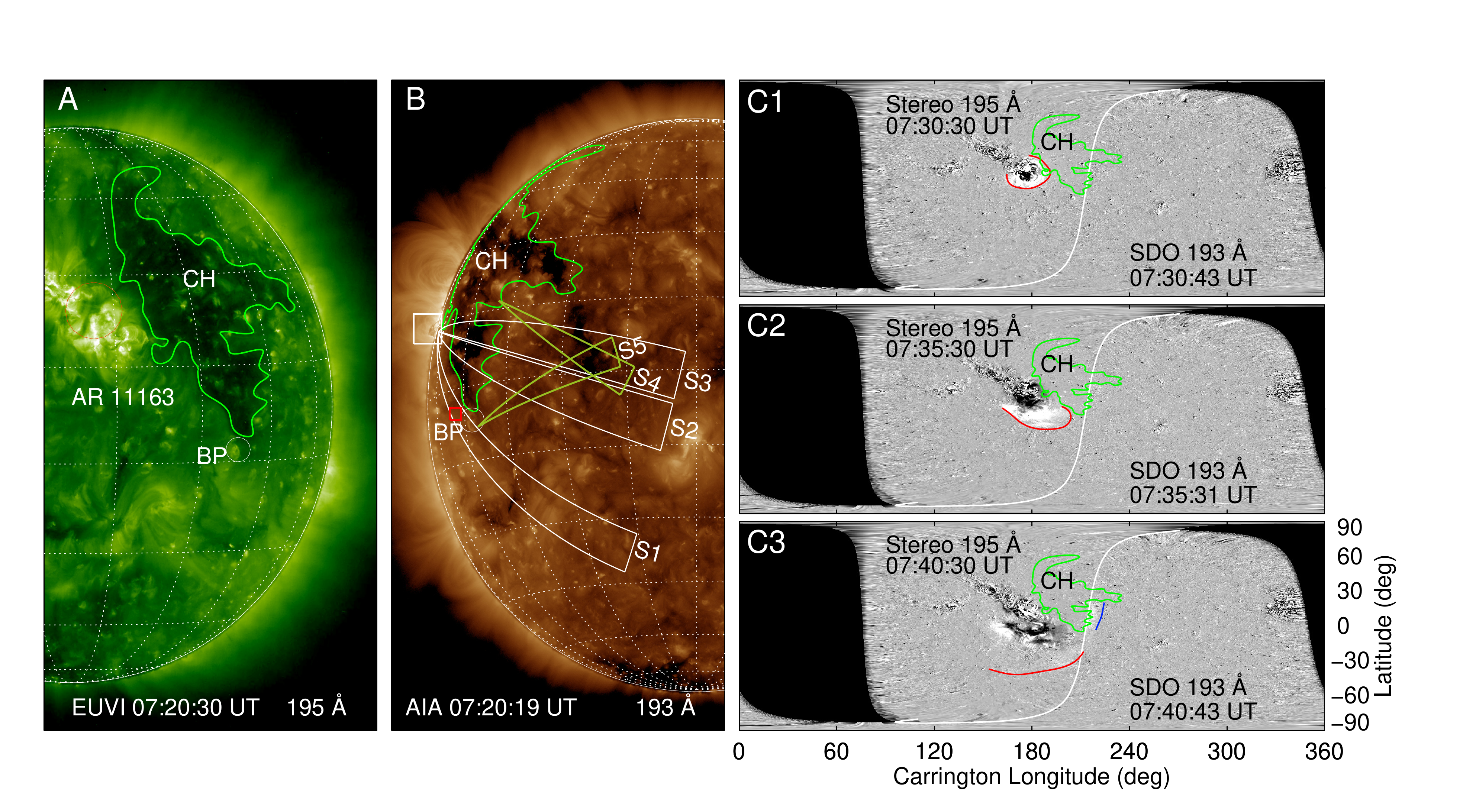}
\caption{STEREO-B/EUVI 195 \AA\ (A) and SDO/AIA 193 \AA\ (B) direct images show the flare and initial coronal condition of the eruption source region. The closed region in panel (A) denoted with “CH” represents the corona hole in the west of the active region in the view of STEREO-B, whose boundary is also projected onto panel (B), panels C1-C3, and \nfig{evolution}. The spherical sectors S1-S5 in panel (B) are used to obtain time-distance stack plots. The circle denoted with “BP” marks the bright point. The white box in panel (B) is used to collect AIA light curves shown in \nfig{flux}, while red box  shows the region used to estimate the variation in density and temperature using differential emission measure in Section 3. The right row shows synoptic maps constructed from running-difference images of  EUVI 195 \AA\ and AIA 193 \AA. The red and blue lines represented the wavefronts propagated in quiet-Sun and transmitted out of the CH, respectively. The white curves indicate the image boundary observed from the two spacecraft, while the black region is unobserved from the two spacecraft. \textcolor{black}{An animation of panels C1-C3 is available. The animation covers 07:01:55 UT\,--\,08:00:19 UT with a 5 minutes cadence. In the animation this sequence appears at the bottom while the AIA 193 \AA\ and 171 \AA\ sequence from Figure 2 is shown at the top. The animation duration is 6 s. (An animation of this figure is available.)}
\label{overview}}
\end{center}
\end{figure*}

\begin{figure*}
\begin{center}
\epsscale{1.}

\plotone{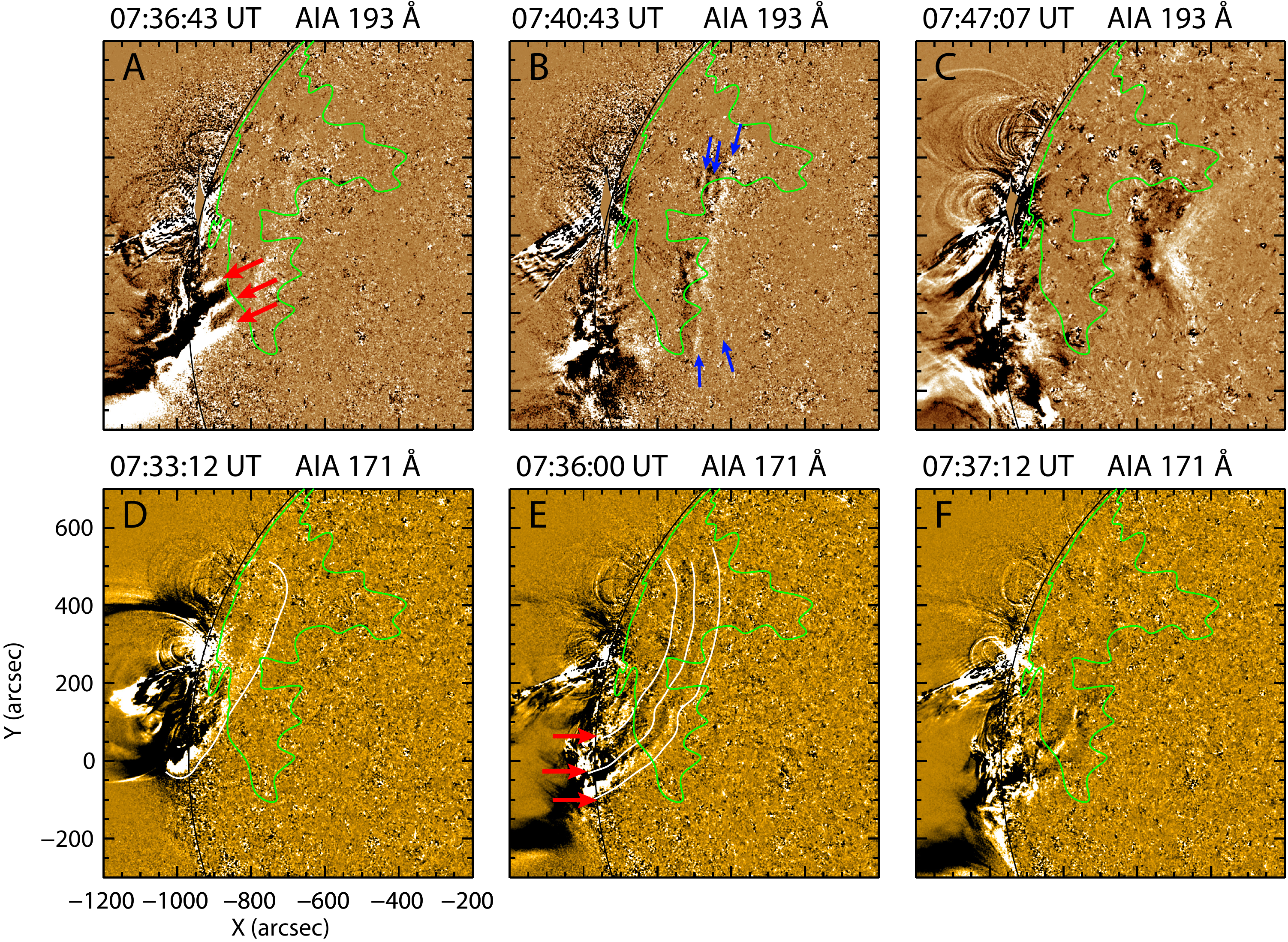}
\caption{AIA 193 \AA\ (A-C) and 171 \AA\ (D-F) running-difference images show the evolution of the wave train. The white curves in panels (D) and (E) tracing the wavefront are added to visualize the wave evolution, which is drawn by connecting a sequence of measurement points. The red and blue arrows point to the wavefronts at the different evolutional stages. The closed region marked the lunate CH boundary. \textcolor{black}{An animation of the evolution of the wave train in 171 \AA, 193 \AA\ and 211 \AA\  (not shown in the figure) is available. This sequence is shown at the top of the animation while the bottom is of the synoptic maps of Figure 1. The animation covers 07:01:55 UT\,--\,08:00:19 UT with a 24 s cadence. The animation duration is 6 s. (An animation of this figure is available.)}
\label{evolution} }
\end{center}
\end{figure*}

\nfig{overview} and the associated animation available in the online journal give an overview of the pre-eruption configuration and the evolutionary process of the wave train. On 2011 February 24, one can see that the eruption source region (NOAA AR11163) was located respectively close to the disk center and on the eastern limb from the viewpoints of the STEREO-B and the SDO (see \nfig{overview} (A) and (B)). The separation angle between the two spacecraft was $\sim$95 degrees. A low latitude, lunate CH can be identified on the west of the eruption source region in the STEREO-B 195 \AA\ images, in which the green curve highlights the boundary of the CH at 07:20:30 UT. 
The boundary of the CH is also outlined in the AIA 193 \AA\ images and the synoptic maps made from STEREO-B 195 \AA\ and AIA 193 \AA\ running difference images (see \nfig{overview} (B) and (C1-C3) and \nfig{evolution}). In \nfig{overview} (B), we can see an isolated small CH located at the west of the large lunate CH. Here, the synoptic maps are obtained by firstly transforming the full-disk images into Carrington coordinates and then constructing the synoptic maps. Note that the cadence of the STEREO-B 195 \AA\ images on 2011 February 24 was 5 minutes; therefore, the cadence of the sequence of synoptic maps used to make the animation is 5 minutes.

At 07:30:43 UT, the wavefront exhibited as a bright area with an angle extent of $\sim$270$^\circ$ surrounding the flare kernel (see the red curve in \nfig{overview} (C1)). Since the eruption source region was very close to the east boundary of the CH, the wavefront began to penetrate into the CH right after its formation, and the westward portion disappeared when it propagated within the CH. At about 07:40:43 UT, the wavefront reappeared to the west of the CH (indicated with a blue curve in \nfig{overview} (C3)). This process suggests the transmission of the wave train through the CH. The southward portion of the wavefront (indicated by the red curve in \nfig{overview} (C2-C3)) showed a free propagation process because no pronounced coronal structures exist in that region. 
This wave-CH interaction is different from previous observations, where the wavefront stopped at the CH boundary and remained stationary for tens of minutes to hours \citep[e.g.,][]{2000ApJ...545..512D}. However, it is similar to the transmission of EUV waves across ARs \citep{2013ApJ...773L..33S}; in the latter case, the wavefront also first disappeared inside ARs but reappeared at the far-side regions outside ARs. Such a transmission through ARs or CHs manifests the true wave nature of EUV waves. 

The high spatio-temporal resolution AIA images showed more details of the wave train than what was observed in the STEREO-B 195 \AA\ images. The detailed evolution and morphological characteristics are mainly displayed using the running-difference images of AIA 193 \AA\ and 171 \AA\ in \nfig{evolution} since the evolutionary processes are similar to other wavelength bands. 
The first wavefront clearly appeared at $\sim$07:30:00 UT, about 7 minutes after the start of the accompanying flare (07:23 UT). Then, multiple wavefronts sequentially appeared following the first one with a similar shape. In the quiet-Sun region southeast to the CH, the wave train can be identified clearly in AIA 193 \AA\ and 171 \AA\ running-difference images (see the red arrows in \nfig{evolution} (A) and (E)). The westward propagating wavefronts can be identified inside the CH but with a small intensity amplitude in AIA 171 Å running-difference images (see \nfig{evolution} (D-E)). However, the simultaneous AIA 193 \AA\ running-difference images did not capture it.  This phenomenon might be caused by the lower temperature of the CH that can not lead to a significant response in high temperature 193 \AA\ images \citep{2020SoPh..295....6S}.

\begin{figure*}
\epsscale{0.95}
\begin{center}
\plotone{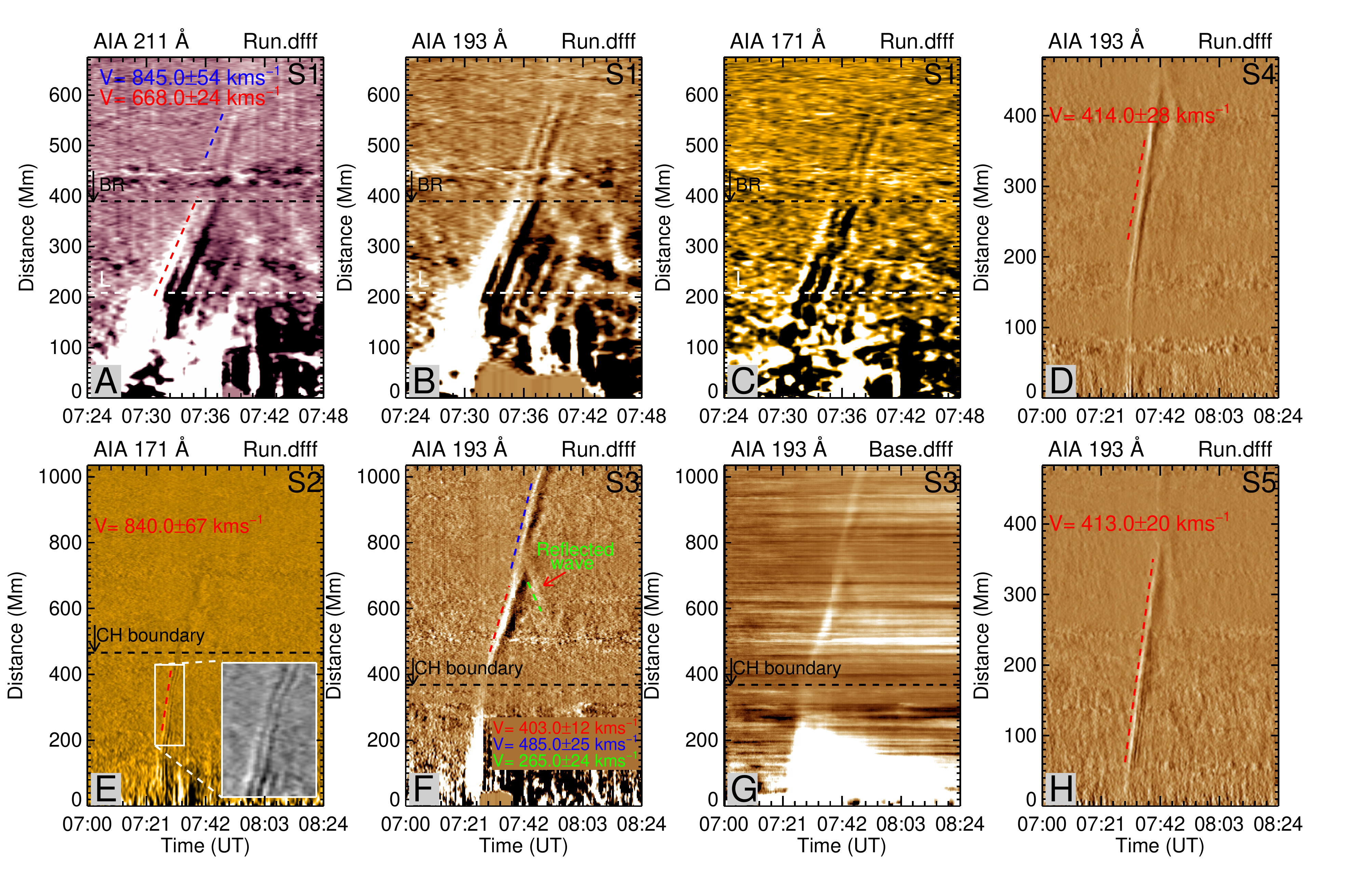}
\caption{Panels (A-C) are time-distance stack plots respectively obtained from AIA 211 \AA, 193 \AA, and 171 \AA\ running-difference images gotten along sector S1 located in the quiet-Sun. The white dashed lines in panels (A-C) point to the positions where the intensity profile is used to analyze the periodicities of the wave train, while the black dashed lines indicate the location of BP. Panels (E-G) are respectively running- (E and F) and base-difference (G) images created along sectors S2 and S3, in which the black dashed lines marked the location of the far-side CH boundary and the inset overlaid in panel (E) is an enlarged view of the wave train inside the CH. Panels (D) and (H) show the evolution of two refracted waves propagated along sectors S4 and S5. The speeds in the different stages are listed in each corresponding panel with different colors.
\label{tdp}}
\end{center}
\end{figure*}

Right after the wave train transmitted through the CH, the wavefront reappeared to the west of the CH. At around 07:40:43 UT, one can observe at least two and three wavefronts close to the south and north arms of the CH’s west boundary (see the blue arrows in \nfig{evolution} (B)). After the transmission, the propagation direction of the transmitted wavefronts showed a significant change: the initial semicircle shape changed to a C-shaped enhanced feature resembling the shape of the west boundary of the CH. The successive refraction should cause a significant change of the propagation direction at the two boundaries of the CH, which acts as a concave lens. Finally, the \textcolor{black}{northern and southern parts of} the transmitted wave train propagated towards and interacted with each other in opposite directions (see \nfig{evolution} (C)). Interestingly, the interaction of the two transmitted wave trains caused a noticeable intensity enhancement at the collision position. This enhancement could be interpreted as the interference effect between the two wave trains because they originate from the same primary wave train, and therefore they should have the same frequency for satisfying the condition of the occurrence of the interference effect. The interference effect will be discussed in \textcolor{black}{detail} in another paper.

\begin{figure*}
\begin{center}
\epsscale{1.}
\plotone{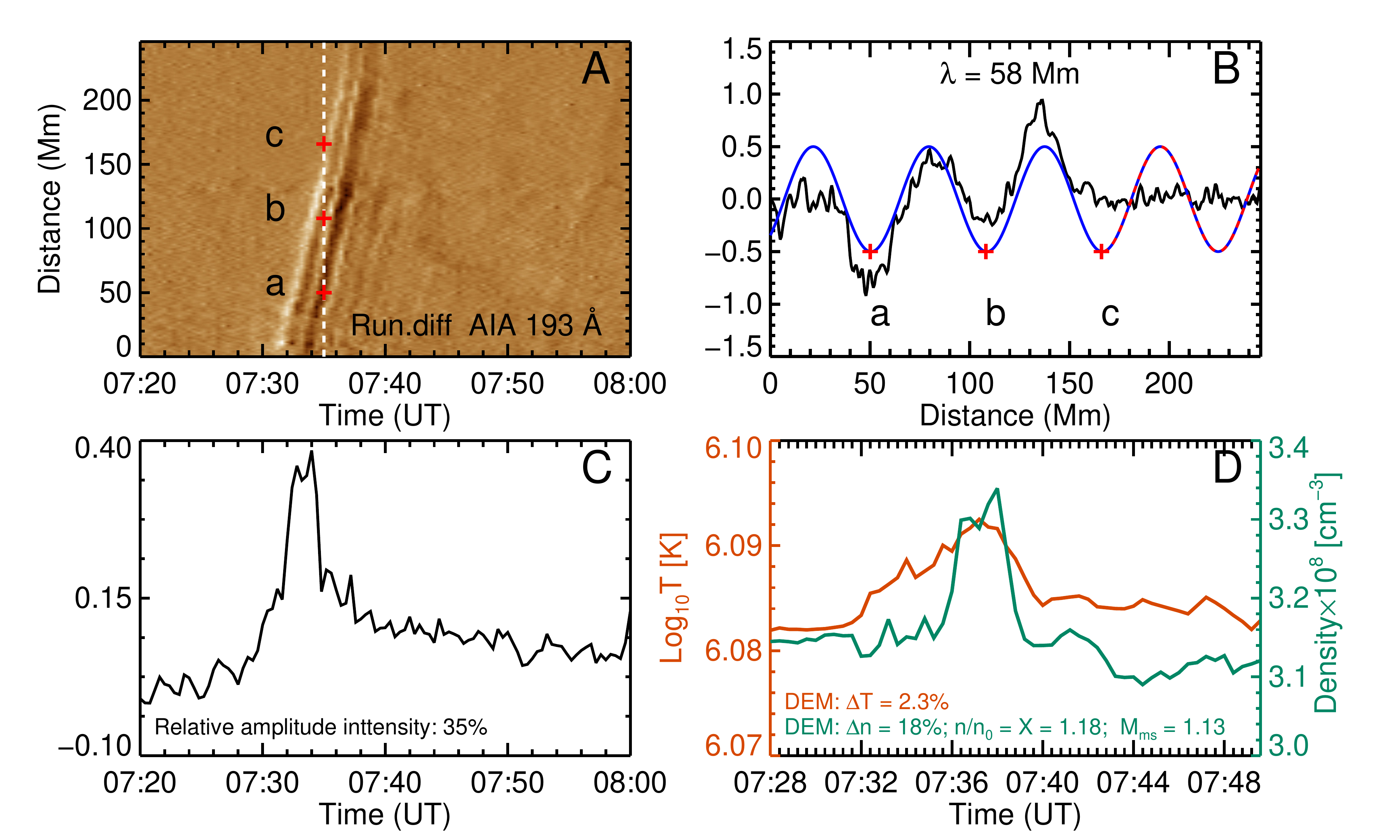}
\caption{Panels (A) shows the running-difference time-distance stack plot of  193 \AA\ along a rectangle slice (not shown here) that tracked the angular bisector of the sector S1. Note that the starting point of the y-axis is not at the flare kernel. Panel (B) shows the profile of the wavefront at 07:35 UT indicated as the vertical white line in panel (A), in which the red crosses, corresponding the location of the red crosses in panel (A), mark the positions of the trough at a distance of 50 Mm, 108 Mm, and 166 Mm, respectively. The blue curve in panel (B) is the result of fitting the wave profile using the harmonic function, which shows the wavelength of 58 Mm. Panel (C) shows the \textcolor{black}{time profile of the relative amplitude} extracted from the position 50 Mm in a base-difference stack plot, \textcolor{black}{as shown in} panel (A). Panel (D) shows the variation in density (dark green) and temperature (orange) estimated in the region outlined by red box in \nfig{overview} (B) using the DEM version developed by \cite{2012A&A...539A.146H}.  The percentage increase in density and temperature due to the wave's passage and the magnetosonic Mach number estimated using Equation 4 are listed in the Figure.
\label{amplitude}}
\end{center}
\end{figure*}

To analyze the kinematics of the wave train, we made the time-distance stack plots using the AIA images along five paths as shown in \nfig{overview} (B), in which sectors S1-S3 originate from the flare kernel, while sectors S4 and S5 are along the propagation directions of the south and north transmitted wave trains, respectively. The time-distance stack plots made from AIA 211\AA, 193 \AA, and 171 {\AA} running-difference images along S1 are plotted in \nfig{tdp} (A-C). Since S1 is located in the quiet-Sun region, the wave train initially propagated freely at an average speed of $\sim$668$\pm$\speed{24}; after the wave train passed through a small bright point (BP) on the path, its speed rapidly increased to more than 845$\pm$\speed{54} (see \nfig{tdp} (A-C)). The time-distance stack plots along sectors S2 and S3 are plotted respectively in \nfig{tdp} (E-G), in which the black dashed line in each panel indicates the west boundary of the CH. In these time-distance stack plots, one can see the significant change of the propagation speed of the wave train at the west boundary of the CH. The speed during the transmission was $\sim$840$\pm$\speed{67} (see \nfig{tdp} (E) and the inset), and it decreased to $\sim$403$\pm$\speed{12} after the transmission (see \nfig{tdp} (E-G)).

To study the kinematics of the two transmitted wave trains, we selected two sectors, S4 and S5, along the propagation directions of the wave trains to obtain the time-distance stack plots. The northward and southward transmitted wave trains propagated with a similar speed of $\sim$\speed{400} as shown in \nfig{tdp} (D) and (H). This value is consistent with the speed of the south portion of the primary wave train propagated in the quiet-Sun region (along sector S3). After the interference of the two transmitted wave trains, they exhibited as a single observable wavefront with a speed of $\sim$485$\pm$\speed{25} (see \nfig{tdp} (F)), slightly higher than those of the transmitted wave trains. At the same time, a reflected wave was observed between $\sim$07:46 UT and 08:00 UT at a speed of about 265$\pm$\speed{24} (see \nfig{tdp} (F) and (G)). The interaction with the small CH on the west of the main CH may cause the origin of this reflected wave. 

To avoid the influence of the amplitude by the different widths of S1 at different distances, we selected a rectangle slice along the angular bisector of S1 to make a new time-distance stack plot to measure the amplitude of the primary wave train in the quiet-Sun region, and the results are shown in \nfig{amplitude} (A). \nfig{amplitude} (B) shows the evolution pattern of the wave train extracted from the running difference time-distance stack plot at 07:35 UT, in which the blue curve is the corresponding fitting result with a harmonic function. The result indicates that the wave train's wavelength $\lambda$ was about 58 Mm. In \nfig{amplitude} (C), we can identify that the relative amplitude intensity is about $35\%$. These parameters are in agreement with that of broad QFP wave trains \citep{2019ApJ...873...22S,2021arXiv211214959S} and typical single pulsed EUV waves \citep{2010ApJ...716L..57V,2015LRSP...12....3W}. However, the intensity amplitude is significantly greater than that of the narrow QFP wave train \citep{2021arXiv211214959S}.

\begin{figure*}
\begin{center}
\epsscale{1.}

\plotone{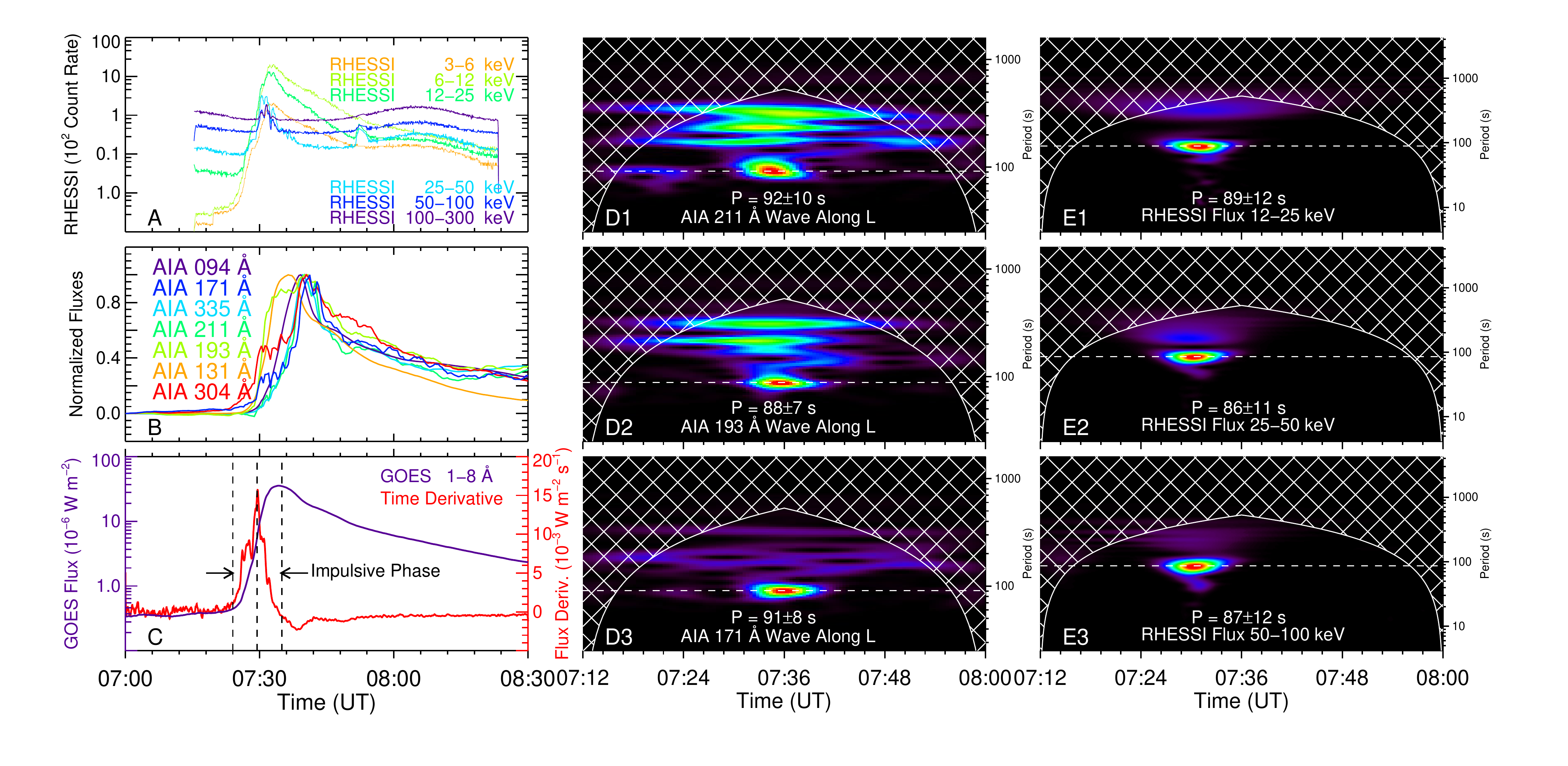}
\caption{Panels (A) and (B) are RHESSI hard X-ray fluxes and normalized light curves within the eruption source region as outlined by the white box in \nfig{overview} (B) measured from different AIA channels, respectively. Panel (C) shows the GOES 1\,--\,8 \AA\ flux (purple) and its derivative curve (red). The middle column (D1-D3) shows the power maps of detrended intensity profiles along the horizontal white dashed lines as shown in \nfig{tdp} (A-C), while the right column (E1-E3) shows the wavelet power maps of RHESSI hard X-ray fluxes in the energy band of 12\,--\,100 keV. In each wavelet power map, the period is highlighted by a white horizontal dashed line, and the corresponding period P is also listed in the figure. 
\label{flux}}
\end{center}
\end{figure*}

To analyze the periodicity of the flare pulsation, we studied the accompanying flare by using the hard and soft X-ray fluxes recorded by the RHESSI and the GOES satellites, and the intensity of light-curves measured from the AIA images around the flaring kernel (see \nfig{flux} (A-C) ). During the impulsive rising phase of the flare (07:24 UT-07:35 UT) as indicated by the vertical dotted lines in \nfig{flux} (C), there are at least four distinct bumps that can be identified in the AIA 171 \AA\ and 304 \AA\ light curves in \nfig{flux} (B). In the meantime, similar bumps can also be observed in the derivative curve of the GOES 1\,--\,8 \AA\ soft X-ray flux (see \nfig{flux} (C)) and the RHESSI 25\,--\,50 keV and 50\,--\,100 keV energy bands (see \nfig{flux} (A)). Generally, the appearance of these bumps may manifest the periodic energy release process in the flare. Detailed estimation suggests that the average period of these bumps was about $\sim$90 seconds. We further analyzed the periodicity of the wave train and the accompanying flare using the wavelet technique \citep{1998BAMS...79...61T} that has been widely used to analyze the periodicity of time-depended one-dimensional data. In our analysis, we choose the ``Morlet'' function as the mother function in the wavelet software, and the results are shown in \nfig{flux} (D1-E3). To analyze the period of the wave train, we extracted the intensity profile along the horizontal white dotted lines marked with L from the time-distance stack plots in \nfig{tdp} (A-C), and the corresponding wavelet power maps are shown in \nfig{flux} (D1-D3). Clearly, the main period of the wave train was about $\sim$90 seconds.
Interestingly, using the relation $\rm v_{ph}=\frac{\lambda}{P}$, we get the phase speed is about \speed{644}, which is consistent with that obtained from the time-distance stack plots as shown in \nfig{tdp} (A-C). This result further implies that the observed EUV waves are non-dispersion in their formation heights corresponding to the AIA channels, similar to the QFP wave confined in the loop system \citep{2011ApJ...736L..13L}. The RHESSI hard X-ray flux curves are used to investigate the period of the flare pulsations since they represent the nonthermal emissions produced by high-energy particles. \nfig{flux} (E1-E3) show the wavelet power maps obtained based on the high energy bands of RHESSI 12\,--\,25 keV, 25\,--\,50 keV, and 50\,--\,100 keV, respectively. It is clear that the period of the flare pulsations was also $\sim$90 seconds, which is in good agreement with that of the observed wave train. 
It should be pointed out that the main periods (errors) of the wave train and the flare were determined by the peak (full width at half maximum) of the corresponding global wavelet power spectrums.  The common 90 seconds periodicities of the wave train and the accompanying flare strongly suggest that the two different phenomena should be originated from the same physical process, such as the nonlinear magnetic reconnection process in the flare \citep{2021arXiv211214959S}. 

\section{Plasma Diagnostics}
{In this section, we study the variations of the plasma temperature and density during the passage of the wave train. The temperature distribution of the contributing plasma in the line of sight is characterized by the Differential Emission Measure (DEM) for optically thin emission lines from plasma in thermodynamic equilibrium. The DEM, defined as
\begin{equation}
{\rm DEM}(T)=n_e^2(T)\frac{{\rm d}h}{{\rm d}T},
\end{equation}
where $n_e$ is the number density dependent on the temperature $T$ along the line of sight. This method enables a direct estimation of the variations of the coronal  density and temperature during the wave's passage. To quantify the variations of temperature and density in the region highlighted with red box in \nfig{overview} (B), the DEM of the plasma observed by SDO/AIA was estimated using the inversion code developed by \cite{2012A&A...539A.146H}. 
The DEM inversion was done between 07:27 UT and 07:50 UT. Following \cite{2012ApJ...761...62C, 2015ApJ...812..173V}, the average temperature and density can be defined as,
\begin{equation}
\bar{T}= { \frac{\int_{T} { \rm DEM}(T)T{\rm d}T}{\int_{T}{\rm DEM}(T){\rm d}T}}
\end{equation}
and 
\begin{equation}
\bar{n}= \sqrt{\frac{\int_{T} {\rm DEM}(T){\rm d}T}{h}},
\end{equation}
respectively, where $h$ is the column height of emitting plasma along the line of sight taken as 90 Mm \citep[cf, ][]{2009ApJ...700L.182P,2021ApJ...921...61L}. As shown in \nfig{amplitude} (D), both the temperature (orange) and the density (dark green) exhibit an increase as a result of the wave's passage, with gains of {2.3\%} and 18\%. These variations in temperature and density are consistent with the report of \cite{2015ApJ...812..173V,2021ApJ...921...61L}. Since the measured intensity of the EUV images is as a function of both the temperature and the density, the small percentage increase in temperature indicates that the measured EUV intensity variation is mainly due to the change of the plasma density rather than the temperature.

Assuming that the observed wave train propagated perpendicular to the direction of the magnetic field \citep[cf,][]{2002A&A...394..299V,2021ApJ...921...61L}; this is reasonable since the magnetic field in the quiet-Sun corona has a strong vertical component. The magnetosonic Mach number $M_{ms}$ can be calculated using,
\begin{equation}
M_{ms}=\sqrt{\frac{X(X+5+5\beta)}{(4-X)(2+5\beta/3)}}
\end{equation}
, where $X$ is the density compression ratio, defined as $X=n/n_0$, and $\beta$ is the plasma-$\beta$ (here taken 0.1 follow \cite{2011ApJ...739...89M}). The $M_{ms}$ is 1.13 taking the density compression ratio of 1.18 estimated by the DEM, suggesting that the observed wave train was weakly shocked. 

\section{Discussions and Conclusions} 
By combining the high spatiotemporal resolution and multiple angles observations taken by SDO and STEREO-B, we studied the generation mechanism and the propagation behaviors of a broad QFP wave train in association with a GOES M3.5 flare and a partial halo CME on 2011 February 24. Based on our analysis results, we propose that the observed QFP wave train was probably driven by the pressure pulses caused by the intermittent energy release in the accompanying flare. In addition, for the first time, we reported the transmission of the wave train through a low latitude CH and the interference effect between the transmitted wave trains. The high projection speed (\speed{668$\pm$24}), Mach number (1.13), transmission phenomenon, and the interference effect of the wave train together suggest that the observed wave train should be a fast-mode magnetosonic wave or a weak shock. In addition, based on the DEM estimation, we find that both the corona's density and the temperature increased after the passage of the wave train, which might indicate the heating of coronal plasma by the wave.

We studied the complete transmission process of the QFP wave train through the CH, although the wave signal was very weak with respect to that in the quiet-Sun. The speed of the wave train during the transmission was $\sim$840$\pm$\speed{67}, which is $\sim$20\% faster than that in the quiet-Sun region. This result is consistent with previous observations and simulations, i.e., the velocity of fast-mode magnetosonic waves propagating inside strong magnetic field strength regions such as CHs and ARs are faster than those in the quiet-Sun region \citep{2009ApJ...691L.123G,2012ApJ...756..143O,2019ApJ...878..106H,2010ApJ...713.1008S}. The faster magnetosonic wave speeds inside CHs are owing to the higher magnetic field strength and lower plasma density inside CHs; such characteristics of CHs can lead to a higher Alfv\'en speed inside CHs than that in the quiet-Sun region. The intensity amplitude of the wave train reduced significantly inside the CH, compared to that in the quiet-Sun region. This result can be interpreted as the result of the conservation of energy \citep{2021ApJ...911..118D}, i.e., as an EUV wave enters a strong magnetic field region (such as a CH) from a weak magnetic field region (such as the quiet-Sun region), the leading part of the wavefront will speeds up, but the trailing part does not, which naturally results in the widening of the perturbation profile and therefore the decrease of the intensity amplitude. In general, the kinetic energy of a wave is directly proportional to the integral of the mass density and the square of the wave amplitude over the whole wave packet. In addition, we find that the wave train had an elevated speed of \speed{845} after its passage through the small BP, as the results reported in \cite{2012ApJ...754....7S} and \cite{ 2019ApJ...878..106H}. The projection effect may cause this since the BP was close to the disk limb or the change of wave's propagation direction due to the refraction caused by the BP. After leaving the west boundary of the CH, the intensity of the wavefront was significantly enhanced at the location where the southwestward and northeastward propagating transmitted wavefronts collided head-on. {We propose that the amplitude enhancement was caused by the interference between the two transmitted wave trains. This explanation is reasonable because the two transmitted wave trains were separated from the same wave train; therefore, they had the same frequency as the primary wave train. These conditions provide the necessary physical premise for the occurrence of the interference effect. We believe the enhancement results from the interference effect. The ideal situation is that \textcolor{black}{  the southern wave train propagates towards the north, while the northern wave train propagates towards the south}, after the interference effect. However, the actual problem is that the waves themselves have become very weak after propagating a long distance, which leads to the observed wave looking like a single wavefront after the collision. In this event, the wave train propagated at fast magnetosonic wave speed, exhibiting refraction, reflection, transmission, and interference effects. These characteristics strongly suggest that the observed wave train should be a fast-mode MHD wave in nature. 
The wave train propagated along the solar surface with an angular extent of about $270^\circ$ and a relative maximum intensity amplitude of about 35\% relative to the unperturbed background corona. These parameters are significantly larger than those QFP wave trains along open or closed coronal loops (i.e., narrow QFP wave trains), where the angular width and the maximum intensity amplitude are in the range of $10^\circ$\,--\,$60^\circ$ and 1\%\,--\,5\%, respectively \citep{2014SoPh..289.3233L,2021arXiv211214959S}. These differences suggest that the observed wave train belongs to the broad type of QFP wave trains as proposed in \cite{2021arXiv211214959S}, which have a relatively larger intensity amplitude, higher energy flux, and larger angular extent than those of narrow QFP wave trains along coronal loops. The magnetosonic Mach number $M_{ms}$ of the observed wave train is about 1.13, which suggests that it should be a non-linear fast-mode magnetosonic wave or a weak shock.

A study on the relationship among flares, CMEs, and waves is essential for diagnosing the generation mechanism of EUV waves. Generally, there are two competing candidate drivers for EUV waves, namely, flares and CMEs. One of the main reasons for the controversy about the origin of coronal waves is the synchronization of the CME acceleration phase and the impulsive phase of the associated flare \citep{2004ApJ...604..420Z,2014ApJ...797...37L}. Therefore it is hard to distinguish whether a particular EUV wave is driven by a CME or ignited by a flare. {Considering 100 Mm was confirmed as a typical distance (i.e., the distance of the earliest observed wavefront from the extrapolated radiant point) for the appearance of a coronal wave \citep{2008ApJ...681L.113V,2011A&A...532A.151W,2012ApJ...752L..23S}, we estimated that the beginning time of the first wavefront was about 07:28 UT, which was derived from extending the red sloped line in \nfig{tdp} (A) down to the distance 100\,Mm. Actually, the beginning time of the first wave should be slightly earlier than 07:28 UT because the eruption source was slightly behind the solar disk from the perspective of the SDO.  
We selected an azimuthal path above the limb to get the information of the CME acceleration phase using the SDO/AIA data. The result indicates that the start time of the CME's acceleration phase was $\sim$07:30 UT, which was behind the beginning time 07:28 UT of the first wave. In contrast, the beginning time of the wave train was $\sim$2 minutes behind the onset of the flare QPPs (07:26 UT, see \nfig{flux}(E1-E3))). This time delay is reasonable for a wave train generated by flare QPPs because the first wave was detected at a region far away from the flare source.  Although these estimated times have large errors, they can still roughly reflect the relationship between CME, flare, and wave in time. Therefore, we believe the wave train should be triggered by the accompanying flare rather than the CME.  }Recent studies also found that coronal jets can also launch large-scale EUV waves directly ahead of the jet top \citep{2018ApJ...861..105S} and indirectly \textcolor{black}{caused} by a sudden expansion of nearby coronal loops through jet-loop interaction \citep{2018ApJ...860L...8S,2018MNRAS.480L..63S}. In the scenario of piston-driven shocks, the piston (CME) can generate a shock wave ahead of the driver, and the wave will freely propagate once it decouples from the CME. However, such EUV waves often show only one wavefront in many observations. Therefore, it is hard to understand how a single CME can produce a wave train with multiple coherent wavefronts. In the line of this thought, we prefer to propose that the present wave train did not drive by the associate CME. On the other hand, the light curves based on the EUV observations of AIA and hard X-ray fluxes in high energy bands (12\,--\,100 keV) based on the RHESSI observations indicate that the period of the wave trains was consistent with that of the flare QPP. This result strongly suggests that the generation of the wave train was probably caused by the intermittent energy release process in the flare.

\textcolor{black}{Flare QPPs is defined as periodic intensity variations of flare light curves with characteristic periods ranging from a fraction of a second to several tens of minutes \citep{2020ApJ...893L..17L,2020ApJ...893....7L,2021ApJ...921..179L}, which has two possible mechanisms: the intermittent energy release/reconnection and MHD oscillations \citep{2019PPCF...61a4024N,2021SSRv..217...34W}.} Observationally, some periods of QFP wave trains are found to be consistent with QPPs, suggesting their common origins.
Many numerical simulations based on magnetic reconnection successfully reproduce the broad QFP wave trains with physical parameters consistent with observations, such as morphology, intensity amplitude, period, and speed. In the simulation of \cite{2015ApJ...800..111Y} based on the interchange reconnection, multiple wave trains were consecutively launched from the outflow region due to the collision between the plasmoids and the field in the outflow region. As the authors mentioned, the simulated wave train propagates isotropically from the source with a speed of \speed{1000}, rather than constrained in funnels with narrow angular extents. Using two-dimensional MHD simulation, \cite{ 2016ApJ...823..150T} revealed that \textcolor{black}{the waves could be spontaneously generated by the oscillations of the strong magnetic field due to quasi-steady impingement of the reconnection outflow}. The exciting process is similar to the sound generated by an externally driven tuning fork. \cite{2021ApJ...911L...8W} reproduced the broad QFP wave train through a three-dimensional radiative MHD simulation. In that simulation, the wave train with dome shape propagated perpendicular to the magnetic field lines with a speed of $\sim$\speed{550\,--\,700}, similar to that of the wave train reported here. The authors proposed that the QFP wave train was possibly driven by the QPP energy release in the accompanying flare. These simulations provided additional evidence that intermitted energy release mechanisms do excite broad QFP wave trains. However, it is necessary to appreciate that the igniting mechanism of the broad QFP wave trains may be diverse and intricate because (i) in the case reported by \cite{ 2019ApJ...873...22S}, whose periods are completely unassociated with the accompanying flares instead consistent with the unwinding of helical structures of filament. (ii) the broad QFP wave trains are also possibly generated by the leaky components of the impulsively generated wave trains \citep{2017ApJ...847L..21P}. The elaborate relationship between the broad QFP wave trains and flares has not been established. More detailed observational and numerical investigations of QFP wave trains are desirable in the future.

\begin{acknowledgments}
We would like to thank the anonymous referee for his/her many valuable suggestions and comments for improving the quality of this article and the data support from the SDO, GOES, and SOHO science teams. This work is supported by the Natural Science Foundation of China (11922307,12173083,11773068,41774179,11633008), the Yunnan Science Foundation for Distinguished Young Scholars (202101AV070004), the National Key R\&D Program of China (2019YFA0405000), the Specialized Research Fund for State Key Laboratories, and the West Light Foundation of Chinese Academy of Sciences.
\end{acknowledgments}

\bibliographystyle{aasjournal}
 \bibliography{euv_wave}

\end{document}